\documentclass[fleqn,12pt]{wlscirep}
\usepackage{wrapfig, framed, caption}
\usepackage[utf8]{inputenc}
\usepackage{heuristica}
\usepackage{graphicx}
\usepackage{wrapfig, framed, caption}
\usepackage{tcolorbox}
\usepackage{xr}
\usepackage{hyperref}
\usepackage{xcolor}

\makeatletter
\newcommand*{\addFileDependency}[1]{
  \typeout{(#1)}
  \@addtofilelist{#1}
  \IfFileExists{#1}{}{\typeout{No file #1.}}
}
\makeatother

\usepackage{setspace}
\usepackage[normalem]{ulem}
\useunder{\uline}{\ulined}{}
\usepackage{xparse}
\newsavebox{\fminipagebox}
\NewDocumentEnvironment{fminipage}{m O{\fboxsep}}
 {\par\kern#2\noindent\begin{lrbox}{\fminipagebox}
  \begin{minipage}{#1}\ignorespaces
 \end{minipage}\end{lrbox}%
  \makebox[#1]{%
    \kern\dimexpr-\fboxsep-\fboxrule\relax
    \fbox{\usebox{\fminipagebox}}%
    \kern\dimexpr-\fboxsep-\fboxrule\relax
  }\par\kern#2
 }

\title{Genius Cliques: Mapping out the Nobel Network

}

\author[*]{Mil\'an Janosov}

\affil[*]{milan@janosov.com, \href{www.janosov.com}{www.janosov.com}}

\begin{document}

\maketitle


\section*{Abstract}
{\small

In this short piece, I delved into the connections of Nobel laureates by applying Network Science methods to and public data collected from Wikipedia. I uncovered the existence of a central "giant component" in the Nobel laureate network, highlighting the core-periphery structure and the disparity in visibility among laureates. I explored the dominance of laureates in the fields of science and humanities, revealing a polarization that contradicts the trend of interdisciplinary research. Furthermore, it the finding sheds light on the underrepresentation of female laureates in certain Nobel Prize categories.

}

\vspace{0.5cm}
{\small {\bf Keywords}: network science, social network analysis, Nobel prize, data science, science of science}

\vspace{1.0cm}
{\it \hspace{-1cm} Published in Nightingale, Journal of the Data Visualization Society, Issue 202, Winter 2022~\cite{nightingale}. }
\vspace{1.0cm}


\section{Inspiring Scientists in and out of the Nobel Circle}

Even though I got my Ph.D. in Network and Data Science, I have always stayed close to my roots, especially in Physics, whenever seeking inspiration. Growing up in Hungary, I was particularly amazed by the achievements of “The Martians,” a group of renowned scientists who emigrated from Hungary to the US around World War II. Interestingly, some of them even went to the same high school.

The Martians included, for instance, Leó Szilárd, who not only discovered the theory of nuclear chain reaction but also co-patented the refrigerator with Albert Einstein and Eugene Wigner–a key scientist at the Manhattan Project–leading the development of the first nuclear reactor. For his contributions, Wigner received the Nobel Prize in Physics in 1963, numbering among the 18 Nobel Prizes that have been linked to thinkers with Hungarian origins.

Those 18 prizes only measure about three percent of all the Nobel Prizes ever awarded. In fact, since 1901, about 600 prizes have gone to somewhat less than a thousand laureates in the fields of Physics, Chemistry, Physiology or Medicine, Literature, Peace, and–starting in 1969–Economics. The site NobelPrize.org highlights other exciting statistics about the prize and its awardees: from the youngest (17 years old) and oldest (97) laureates to multiple-prize winners such as John Bardeen (Physics, 1956 and 1972), Linus Pauling (Chemistry 1954, Peace 1962), and Marie Skłodowska-Curie (Physics 1903, Chemistry 1911).

The Curie family dominated the Nobel. Marie Curie first shared a prize with her husband, Pierre, and later received a second award. Additionally, the mighty couple produced a Nobel-winning heir. Their daughter, Irène Curie, who shared the recognition with her husband, Frédéric Joliot, was awarded the prize in the field of Chemistry in 1935. Marie Curie was a member of another fabulous example of the interlinked small world of laureates (sadly, Pierre passed away in 1906): the Solvay Conference on Physics in 1911. It was probably the most impressive line-up in science ever: 27 of the 29 participants had either already won, or later received, the Nobel Prize.


\section{Building the Nobel Network}

The stories of the Martians, the Curie family, the Manhattan Project, and the Solvay Conference all suggest that, behind the scenes, some seriously intertwined social networks are at work among Nobel laureates. To trace this network~\cite{netsci, sos}, I went to the most widely used online encyclopedia, Wikipedia, and collected the Wiki page text of each laureate.

\begin{figure}[!hbt]
\centering
\includegraphics[width=0.95\textwidth]{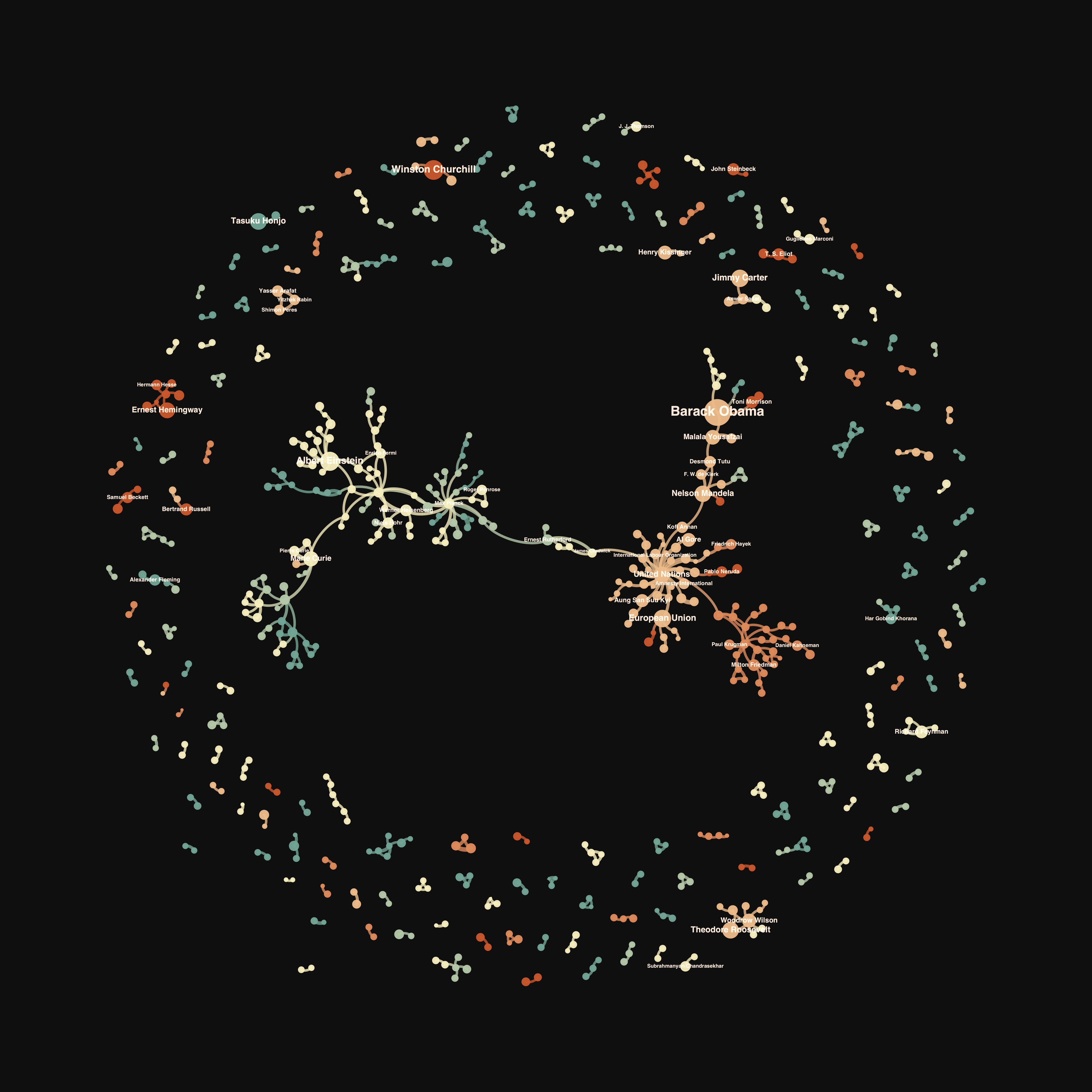}
\caption{Nobel Network. The network of Nobel laureates with at least one connection, based on the cross-references between their Wikipedia pages. Each node corresponds to a laurate, edge widths measure the number of cross references, and node size is proportional to the total view count of their Wiki pages. Color encodes the disciplines they were awarded (in the case of multiple different awards, a color was picked at random from the awarded disciplines). Nodes with the highest view counts are labeled.}
\label{fig:fig1}
\end{figure}

Then, in each laureate’s page text, I counted the mentions of all the other names, noting whether any pairings shared a common history noteworthy for Wikipedia. This way, I built a network of 682 nodes and 588 links, where nodes correspond to laureates, and the strength of the link between two nodes is proportional to the number of times their Wiki sites reference each other. Additionally, I downloaded the total view count of each laureate’s page and set their network node size proportional to the logarithm of that number. This node scaling eventually highlighted those that have become household names. To finalize the network visualization, I applied color coding that corresponds to the scientific disciplines. You may find the result in Figure \ref{fig:fig1}.

\section{The Nobel Network’s Lessons}

To me, as a network scientist, the first and most striking observation about the network is its core-periphery separation: a large, connected component in the center (a so-called giant component) which contains more than 30 percent of the nodes, and a fragmented ring around it with smaller network components, with sizes up to ten nodes. The most frequent component sizes are as few as two and three nodes, which aligns well with the fact that the Nobel Prize can be shared among a maximum of three laureates, and shared prizes are becoming more and more common in the majority of fields.

I also realized that nodes in the giant component are larger, meaning significantly higher visibility and a greater number of search hits for those laureates, as measured by logarithm of their Wiki view counts. After looking into the data, it turns out that the median Wiki-view count is 351,005 in the central component, while only 170,510 for the outer ring, and the average view count value is about 2.5 times higher for the central component than for the outer ring. So it seems, the central clique is way more popular!

\begin{figure}[!hbt]
\centering
\includegraphics[width=0.75\textwidth]{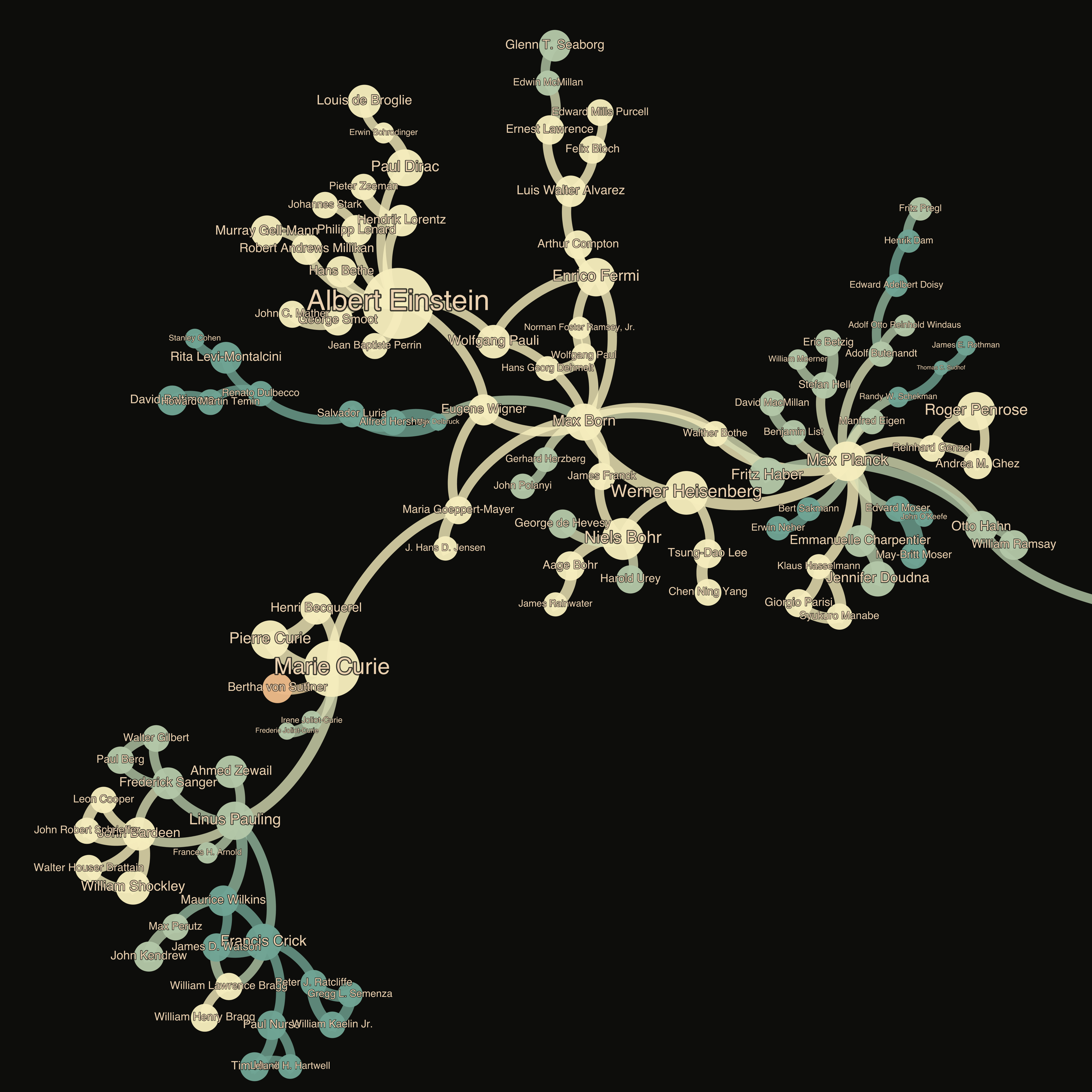}
\caption{Zoom-in of Figure 1, focusing on the clique in sciences.}
\label{fig:fig2}
\end{figure}

But who are they? The coloring with the green-yellowish shades versus reddish tones is meant to distinguish sciences from humanities, coinciding with the left and right sides of the giant component. These sides are linked by Sir James Chadwick, who won the 1935 Nobel Prize in Physics for discovering the neutron and who also became a scientific advisor to the United Nations. The science side (Figure \ref{fig:fig2}), headlining researchers like Albert Einstein and Max Planck, seems to have a strong root in the Prussian Academy of Sciences (1700–1945) and is also strong amongst the founders of modern Physics, from the Curies to Enrico Fermi and Eugene Wigner or György Hevesy (both with Hungarian and Martian roots).

\begin{figure}[!hbt]
\centering
\includegraphics[width=0.75\textwidth]{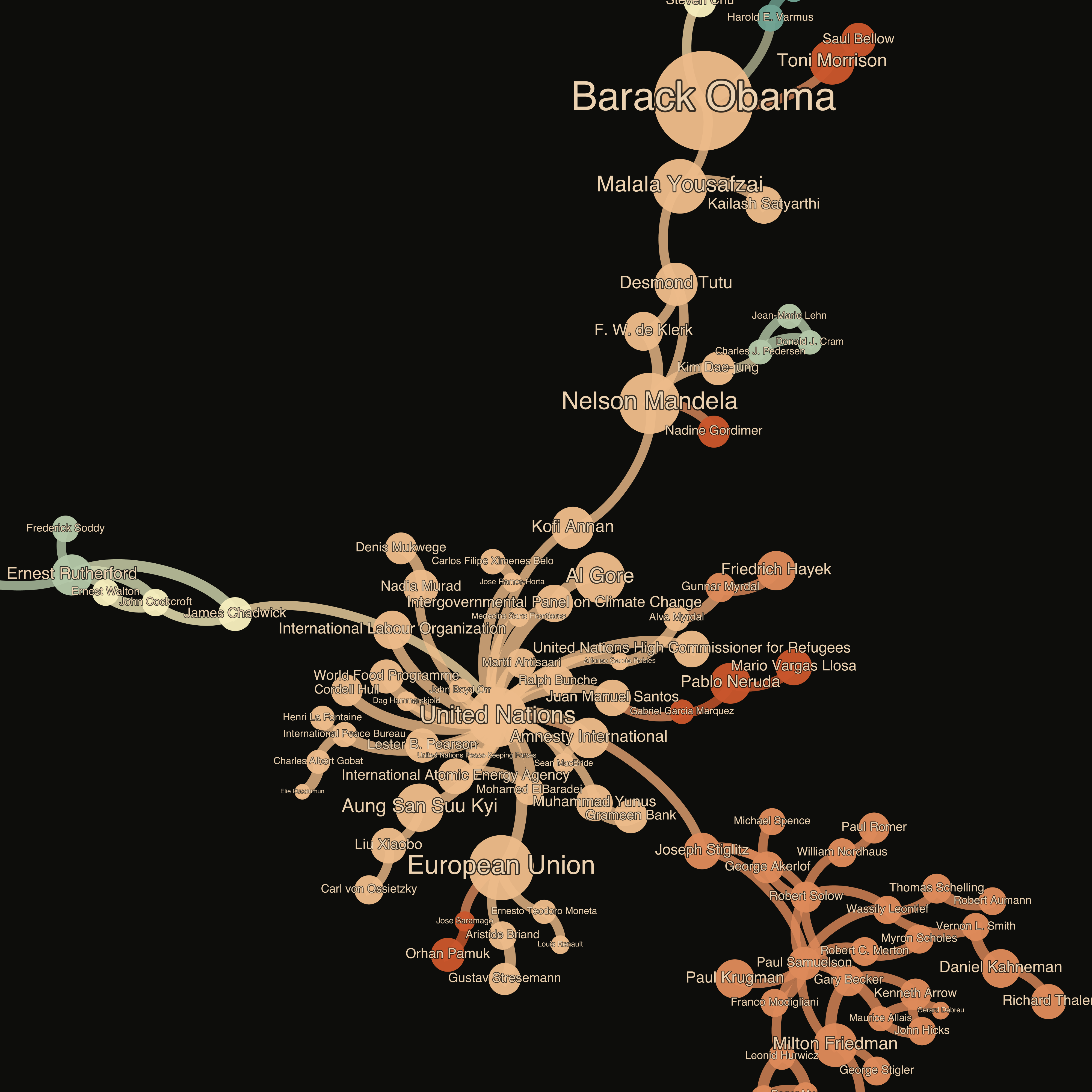}
\caption{Zoom-in of Figure 1, focusing on the clique in humanities.}
\label{fig:fig3}
\end{figure}

On the humanities side (Figure \ref{fig:fig3}), we can see some quite popular figures. Apparently, science is not the way to world fame! There are immediately two central laureate organizations that strike the eye: the European Union and the United Nations, both awarded Nobel Peace Prizes. Notable individuals include prominent politicians, such as Barack Obama or Henry Kissinger, the human rights activist Nelson Mandela, and the economist Milton Friedman (with Hungarian, but non-Martian, roots).

As for the outer parts, there are a few famously social individuals, such as Ernest Hemingway, Winston Churchill, Franklin D. Roosevelt, and Richard Feynman — personally, my favorite Nobel laureate for both his scientific contribution and his playful and eccentric personality. These individuals, despite living busy lives, are somewhat isolated from the network, likely due to the time and geographical locations of their active years compared to other laureates. Additionally, the data may be incomplete here as Wikipedia is neither perfect nor 100 percent accurate in documenting social connections, and sadly (or not?), Facebook did not exist at that time.

Finally, the Hungarians and the Martians: looking at the data, it turns out that many of them are not connected to even a single Nobel laureate, and those who are members of the network are simply scattered around. The reasons behind this are unclear — maybe the legend of the Martians is overrated, or maybe there weren’t enough of them awarded a Nobel to appear in the network visibility. One thing is for sure, though: the Manhattan Project counted seven Nobel laureates while it was operational and, later, a dozen more, but among them only Wigner was from the Martians.

\section{Conclusion}

As inspirational as it is to scan all these names and connections in The Nobel Network, and despite how it makes me truly feel as if I’m “standing on the shoulders of giants,” the network has its flaws. Besides the peripheral Eastern Europeans, we see an elite club emerging in the center with the majority of popular names clustered together in the giant component, excluding two-thirds of the network. This suggests that two-thirds of the laureates just walk away with the prize and go back to their work, and only the remaining third engage in visible connections, be it friendships or collaborations. As “The whole is greater than the sum of the parts,” missing more than 60 percent of those brilliant minds from the central flow of ideas seems a pity.

Even more missed opportunities arise. The central component itself is split into two camps: science and humanities. This polarization very much goes against today’s main direction, interdisciplinary research, which gives us the power to tackle major societal problems never experienced before. Additionally, the network reveals the low number of female laureates. Despite the exceptional history of Marie Curie, only about six percent of laureates were female, most of whom were awarded the Peace Prize (16.5 percent of 109 awarded) and the least of whom earned awards in Physics (1.8 percent of 219 awarded).

Still, all is not lost. Mapping exercises like this one can help reveal these issues, which otherwise are barely visible, even to the most avid Nobel fans. Zooming out and utilizing network science can highlight otherwise hidden patterns and enable understanding, which is the first step in identifying future solutions, be it about gender gaps or elitist cliques.

\section{Disclaimer}

During the process of creating this text, several AI tools were used. In particulary, Grammarly was used for corrections and copy editing throughtout the txt, while ChatGPT 3.5 was used to create the abstract.


\end{document}